\documentclass[12pt]{iopart}
\usepackage{color}
\usepackage{amssymb}
\usepackage{cite}
\begin{document}

\bibliographystyle{prsty}
\input{psfig} 

\letter{Coherently tunable third-order nonlinearity in a nanojunction}
 
\author{Vadim A. Markel\footnote[3]{e-mail: vmarkel@mail.med.upenn.edu}}

\address{Departments of Radiology and Bioengineering, University
  of Pennsylvania, Philadelphia, PA 19104\\ vmarkel@mail.med.upenn.edu}

\date{\today} 

\begin{abstract}
  A possibility of tuning the phase of the third-order Kerr-type
  nonlinear susceptibility in a system consisting of two interacting
  metal nanospheres and a nonlinearly polarizable molecule is
  investigated theoretically and numerically. It is shown that by
  varying the relative inter-sphere separation, it is possible to tune
  the phase of the effective nonlinear susceptibility
  $\chi^{(3)}(\omega;\omega,\omega,-\omega)$ in the whole range from
  $0$ to $2\pi$.
\end{abstract}

\submitto{\JPA}
\maketitle

Recent dramatic advances in nanofabrication made it possible to
design, arrange and assemble nanoparticles with great precision.
Optical and, more generally, electromagnetic properties of
nanostructures have been of great interest in the past
decade~\cite{henneberger_book_93,kreibig_book_95,gaponenko_book_98,shalaev_book_02,kelly_03_1}.
In particular, physical effects due to giant local field enhancement
are subject of active current
research~\cite{kelly_03_1,li-kuiru_03_1,bliokh_04_1,zou_05_1}. In this
respect, nanoparticles of noble metals, especially silver, proved to
be very useful. The remarkable optical propertied on silver
nanostructures are explained by the strong, resonant interaction with
electromagnetic fields in the visible and near-IR spectral range and
by very small Ohmic losses.

The strong enhancement of local fields in small spatial areas is the
consequence of two factors: the heterogeneity of a nanostructure on a
subwavelength scale and the resonant character of interaction of the
electromagnetic field with the nanostructure. Both features are, in
principle, present even in the case of a single isolated nanosphere.
However, the effect becomes much stronger in aggregated nanospheres
due to the effect of {\it plasmon hybridization}~\cite{nordlander_04_1}.
In this case, amplification of local field can become sufficiently
large to make possible detection of Raman radiation from single
molecules, as was demonstrated experimentally in~\cite{kneipp_97_1}.
The Raman enhancement factor $(\vert {\bf E} \vert / \vert {\bf E}_0
\vert )^4$ in the center of a junction between two nanospheres (${\bf
  E}$ and ${\bf E}_0$ - the local and the external fields in the
junction, respectively) was calculated to be $5.5\cdot 10^9$ for a
$1{\rm nm}$ gap between two silver spheres of $60{\rm nm}$ radius each
at $\lambda=497{\rm nm}$~\cite{jiang_j_03_1}. Even larger enhancement,
up to $10^{13}$, was predicted in the so-called nanolens - a linear
chain of several nanospheres of different size~\cite{li-kuiru_03_1}.

The primary focus of research in single-molecule spectroscopy has been
on non-coherent optical processes such as Raman scattering. In this
paper I consider a coherent nonlinear effect, namely, degenerate third
order nonlinearity, and demonstrate that, by changing the geometry of
a nanostructure, it is possible to control not only the amplitude of
the nonlinear response but also its phase (relative to the phase of
the incident field). The nonlinear Kerr effect described by the
third-order susceptibility $\chi^{(3)}(\omega;\omega,\omega,-\omega)$
is responsible for corrections to absorption and refraction. The
possibility to control the phase and tensor structure of
$\chi^{(3)}(\omega;\omega,\omega,-\omega)$ can have numerous
applications, for example, in quantum nondemolition measurements via
the optical Kerr effect~\cite{scully_book_97}.

\begin{figure}
\psfig{file=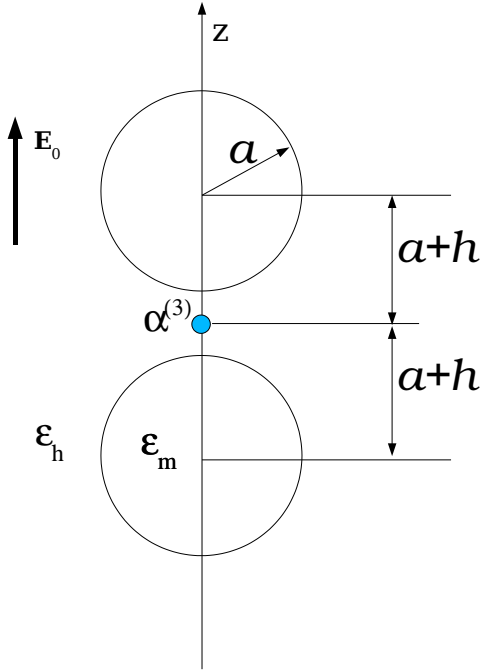,width=8.2cm}
\caption{\label{fig:sketch} Schematic illustration of the physical
  system considered in this paper.}
\end{figure}

Consider a simple physical system shown schematically in
Fig.~\ref{fig:sketch}. Here a nonlinearly-polarizable molecule is
placed in the center of symmetry of two spheres. The radius of each
sphere is denoted by $a$ and the width of the gap (surface-to surface)
by $2h$. Thus, for example, if $a=50{\rm nm}$ and $2h=1{\rm nm}$,
which is the smallest physical gap considered in
Ref.~\cite{jiang_j_03_1}, we have $h/a=0.01$. The system is excited by
an external field with some frequency $\omega$ and wavelength in
vacuum $\lambda=2\pi c/\omega$, linearly polarized along the axis of
symmetry. The latter coincides with the $z$-axis. The whole system is
assumed to be sufficiently small compared to $\lambda$ and we work in
the quasistatic approximation. Further, we use the Drude formula for
the dielectric function of metal, namely,

\begin{equation}
\label{Drude}
\varepsilon = \varepsilon_0 - \omega_p^2/\omega(\omega + i\gamma) \ .
\end{equation}

\noindent
The following parameters are used in the simulations:
$\varepsilon_0=5$, $\gamma/\omega_p=0.002$, and the value of
$\omega_p$ is unspecified. Note that the dielectric function of silver
in the anomalous dispersion region is well described by choosing
$\omega_p \approx 4.6 {\rm sec}^{-1}$ ($\lambda_p \approx 136{\rm
  nm}$).  Finally, we assume that the metal nanoparticles are embedded
in a transparent host medium with the refraction index $n_h=2$
($\varepsilon_h=4$).

Because of the axial symmetry, the dipole moment induced in the
molecule is parallel to the $z$-axis. The third-order nonlinear
correction to the dipole moment oscillating at the same temporal
frequency as the incident field is given by

\begin{equation}
\label{d_nl_def}
d_z^{(NL)}(t) = \alpha^{(3)} E_z \vert E_z \vert^2 = \alpha_{\rm eff}^{(3)}
E_0 \vert E_0 \vert ^2 \exp(-i\omega t)\ .
\end{equation}

\noindent
Here $E_z$ is the amplitude of local electric field at the location of
the molecule and $E_0$ is the amplitude of the external field. Note
that, even if we assume for simplicity that $E_0$ is purely real,
$E_z$ can be complex. In general, there can be an arbitrary phase
shift between the local and the external fields.  The effective
nonlinear polarizability $\alpha_{\rm eff}^{(3)}$ is related to
$\alpha^{(3)}$ by

\begin{equation}
\label{alpha_eff_def}
\alpha_{\rm eff}^{(3)} = G \alpha^{(3)} \ ,
\end{equation}

\noindent
where the enhancement factor $G$ is given by 

\begin{equation}
\label{G_def}
G = \frac{E_z \vert E_z \vert^2 }{E_0 \vert E_0\vert^2} \ .
\end{equation}

\noindent
Since $G$ is, in general, complex, it can influence not only the
magnitude but also the phase of the effective third-order
polarizability. Below, we calculate $G$ numerically and show that its
phase can be varied in its whole range by changing the inter-sphere
separation.

To calculate the local field in the gap, $E_z$, we expand the
polarization inside each sphere in the quasistatic vector spherical
harmonics ${\bf X}_{ilm}^{(1)}({\bf r}) = (la)^{-1/2}
\nabla \psi_{lm}^{(1)}({\bf r} - {\bf r}_i)$. Here $i=1,2$ indexes the
nanospheres, ${\bf r}_i$ are the radius-vectors of the spheres'
centers, $\psi_{lm}^{(1)}({\bf r}) = (r/a)^l Y_{lm}(\hat{\bf r})$, and
$Y_{lm}(\hat{\bf r})$ are spherical functions of the polar angles of
the unit vector $\hat{\bf r}$. Polarization inside each sphere can be
written as

\begin{equation}
\label{P_exp}
{\bf P}({\bf r}) = \sum_{lm} C_{ilm} {\bf X}_{ilm}^{(1)}({\bf r)} \ ,
\ \ {\rm if} \ \vert {\bf r} - {\bf r}_i \vert < a \ ,
\end{equation}

\noindent
where the unknown coefficients $C_{ilm}$ must be found from the
standard boundary conditions applied at the surface of each sphere, or
alternatively, from the integral equation formalism as described
in~\cite{markel_04_3}. From general considerations, it is clear that
$C_{ilm}$ obey a system of linear equations which, in the quasistatic
limit, was obtained in~\cite{gerardy_80_1} and simplified
in~\cite{mackowski_95_1}. In general, this
set of equations has the form

\begin{equation}
\label{lin_syst}
(1/\chi - W)\vert C \rangle = \vert E \rangle \ ,
\end{equation}

\noindent
where $\chi=(3/4\pi)[(\varepsilon - \varepsilon_h)/(\varepsilon +
2\varepsilon_h)]$ is the coupling constant, $W$ the electromagnetic
interaction matrix, and $\vert E \rangle$ is the appropriate
right-hand side defined by the external field. In the case of axial
symmetry, only modes with $m=0$ are excited, so that
$C_{ilm}=C_{il0}\delta_{m0}$. The matrix elements of $W$ needed to
find the solution are

\begin{eqnarray}
\langle il0 \vert W \vert i^{\prime}l^{\prime}0 \rangle = &&  {{l\delta_{l l^{\prime}}\delta_{i i^{\prime}}} \over {2l+1}} +
(1-\delta_{i i^{\prime}})(-1)^{l^{\prime}}[{\rm
  sgn}(z_{i}-z_{i^{\prime}})]^{l+l^{\prime}} 
\nonumber \\ && \times 
\sqrt{{l l^{\prime}} \over
  {(2l+1)(2l^{\prime}+1)}} {{(l+l^{\prime})!} \over
  { l! l^{\prime}! (1+h/a)^{l+l^{\prime}+1}}}
\label{W_def}
\end{eqnarray}

\noindent
and the components of the right-hand side vector in (\ref{lin_syst})
are given by

\begin{equation}
\label{E_il}
\langle i l 0 \vert E \rangle =  E_0 \sqrt{ 4\pi a^3 /3} \ .
\end{equation}

In general, once the coefficients $C_{ilm}$ are found, the scattered
field at an arbitrary point ${\bf r}$ in the host medium can be found
from

\begin{equation}
\label{E_s_C}
{\bf E}_s({\bf r}) = - \sum_{ilm} C_{ilm} \frac{4\pi(l-1)}{3(2l+1)} {\bf
  X}^{(2)}_{ilm}({\bf r}) \ ,
\end{equation}

\noindent
where ${\bf X}^{(2)}_{ilm}({\bf r})=[(l+1)a]^{-1/2}\nabla
\psi^{(2)}_{lm}({\bf r}-{\bf r}_i)$ are the quasistatic vector
spherical harmonics of the second kind and $\psi^{(2)}_{lm}({\bf
  r})=(a/r)^{l+1}Y_{lm}({\bf r})$. For the particular problem
considered in this paper, summation (\ref{E_s_C}) contains only terms
with $m=0$. Further simplification is obtained if the electric field
is evaluated on the axis of symmetry, in which case

\begin{equation}
\label{E_s_f}
{\bf E}_{sz}(z) = E_0 \sqrt{{4\pi} \over {a^3}} \sum_n \frac{\langle E \vert P_n \rangle} {1/\chi - w_n} 
f_n\left( {z\over h} \right) \ ,
\end{equation}

\noindent
where 

\begin{eqnarray}
\label{f_n}
\fl f_n(x) = \sum_{l=1}^{\infty} && (l+1)\sqrt{l \over {2l+1}} \left \{
\frac{\langle 1 l 0 \vert P_n \rangle} {\left[1 + (h/a)(1+x)
  \right]^{l+2} } 
%\right. \nonumber \\ &&  \left. 
- (-1)^l 
\frac{\langle 2 l 0 \vert P_n \rangle} {\left[1 + (h/a)(1-x)
  \right]^{l+2} } \right \} \
\end{eqnarray}

\noindent
and $\vert P_n\rangle$ are the eigenvectors of $W$ with corresponding
eigenvalues $w_n$. Here $E_{sz}$ is the $z$-component of the scattered
field on the axis of symmetry (the $x$- and $y$-components are zero).
The total local field $E_z$ is a superposition of the incident and
scattered fields:

\begin{equation}
\label{e_z}
E_z = E_0 + E_{sz} \ .
\end{equation}

\noindent
Note that we have used the spectral approach to solving
(\ref{lin_syst}). In other words, instead of directly inverting
$1/\chi - W$, we seek eigenvectors and eigenvalues of $W$ and then
obtain the solution in terms of these quantities for an arbitrary
coupling constant $\chi$.

The matrix $W$ is of infinite size and in practical calculations must
be truncated. The truncation order $l_{\rm max}$ required to obtain an
accurate solution depends on the inter-sphere separation. Although,
for any separation, there exist an infinite number of modes, most of
them are antisymmetric~\cite{markel_95_1,markel_97_1} or,
equivalently, dark~\cite{stockman_01_1}. A dark mode is not coupled to
the homogeneous external field because the scalar product $\langle E
\vert P_n \rangle$ is either exactly zero or very small.
Correspondingly, the input of a dark mode to sum (\ref{E_s_f}) is
negligible. Modes which are not dark are referred to as
luminous~\cite{stockman_01_1}.  For a finite value of $h$, there is a
finite number of luminous modes and the spectrum of eigenvalues $w_n$
which correspond to these modes is discrete.  However, as $h$
decreases, the number of luminous modes grows and the intervals
between consecutive values of corresponding eigenvalues $w_n$ approach
zero. When the two spheres touch, the spectrum becomes continuous and,
strictly speaking, can not be adequately described at any finite
truncation order. However, for practical purposes, the matrix still
can be truncated, as long as the resultant discrete density of states
approximates the true continuous function with sufficient precision.
The latter condition depends on the relaxation in the system and is
very difficult to satisfy for silver in the near-IR spectral region
due to the very small non-radiative relaxation. In the simulations
presented below, the minimum ratio $h/a$ is equal to $0.01$. In this
case, all luminous modes are obtained with very high precision at
relatively modest truncation orders. The results reported below were
obtained at $l_{\rm max}=800$ and convergence with machine
accuracy was verified by doubling this number.

\begin{figure}
  \centerline{\footnotesize \input{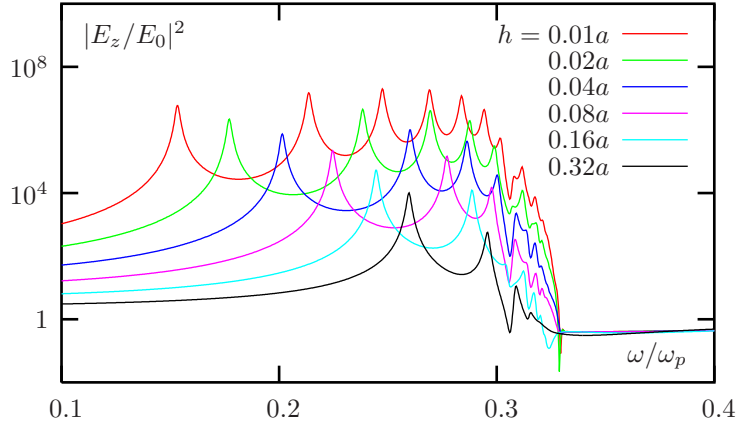}} 
%-------- Uncomment next line for black-and-white version fo the figure
%  \centerline{\footnotesize \input{eo_black_and_white.tex}} 
\caption{\label{fig:eo} The ratio $\vert E_z/E_0 \vert^2$ in the
  center of the inter-sphere gap as a function of $\omega/\omega_p$
  for different relative inter-sphere separations.}
\end{figure}

Now we turn to the numerical results. First, in Fig.~\ref{fig:eo}, we
plot the spectral dependence of the factor $\vert E_z/E_0 \vert^2$ in
the center of the inter-sphere gap for different relative separations
$h/a$. It can be seen that a ``resonance band'' exists in the spectral
region whose bounds depend on the ratio $h/a$. For $h=0.01a$,
resonance interaction takes place for $0.15 \lesssim \omega/\omega_p
\lesssim 0.33$. For $h=0.32a$, the resonance band is smaller, $0.25
\lesssim \omega/\omega_p \lesssim 0.32$. We will be interested in the
frequencies which lie in the resonance band for the smallest value of
$h$ considered, namely, $h=0.01a$.

\begin{figure}
\centerline{\psfig{file=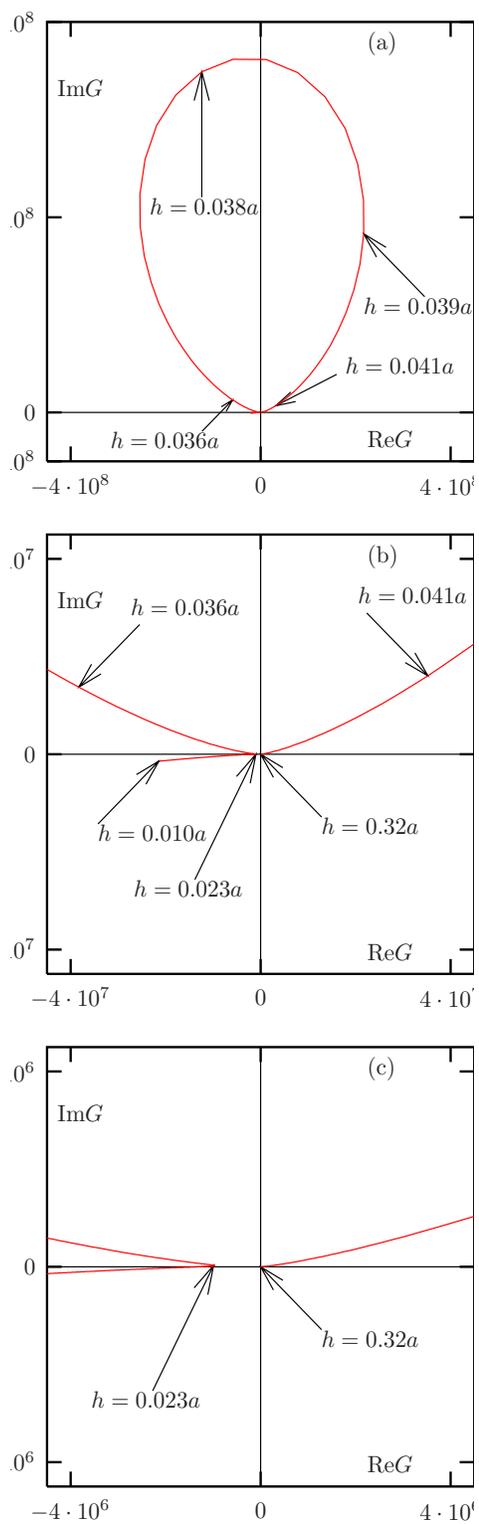,height=22cm,bbllx=200bp,bblly=0bp,bburx=400bp,bbury=700bp,clip=}}
%\centerline{\footnotesize \hspace*{-1.5cm}\input{eh_o=0.20_a.tex}}
%\centerline{\footnotesize \hspace*{-1.5cm}\input{eh_o=0.20_b.tex}}
%\centerline{\footnotesize \hspace*{-1.5cm}\input{eh_o=0.20_c.tex}}
\caption{\label{fig:eh_o=0.20}Parametric plot of the complex
  enhancement factor $G$ as a function of $h/a$ for
  $\omega/\omega_p=0.2$. Graphs (a)-(c) show the same curve on
  different scales.} 
\end{figure}

\begin{figure}
\centerline{\psfig{file=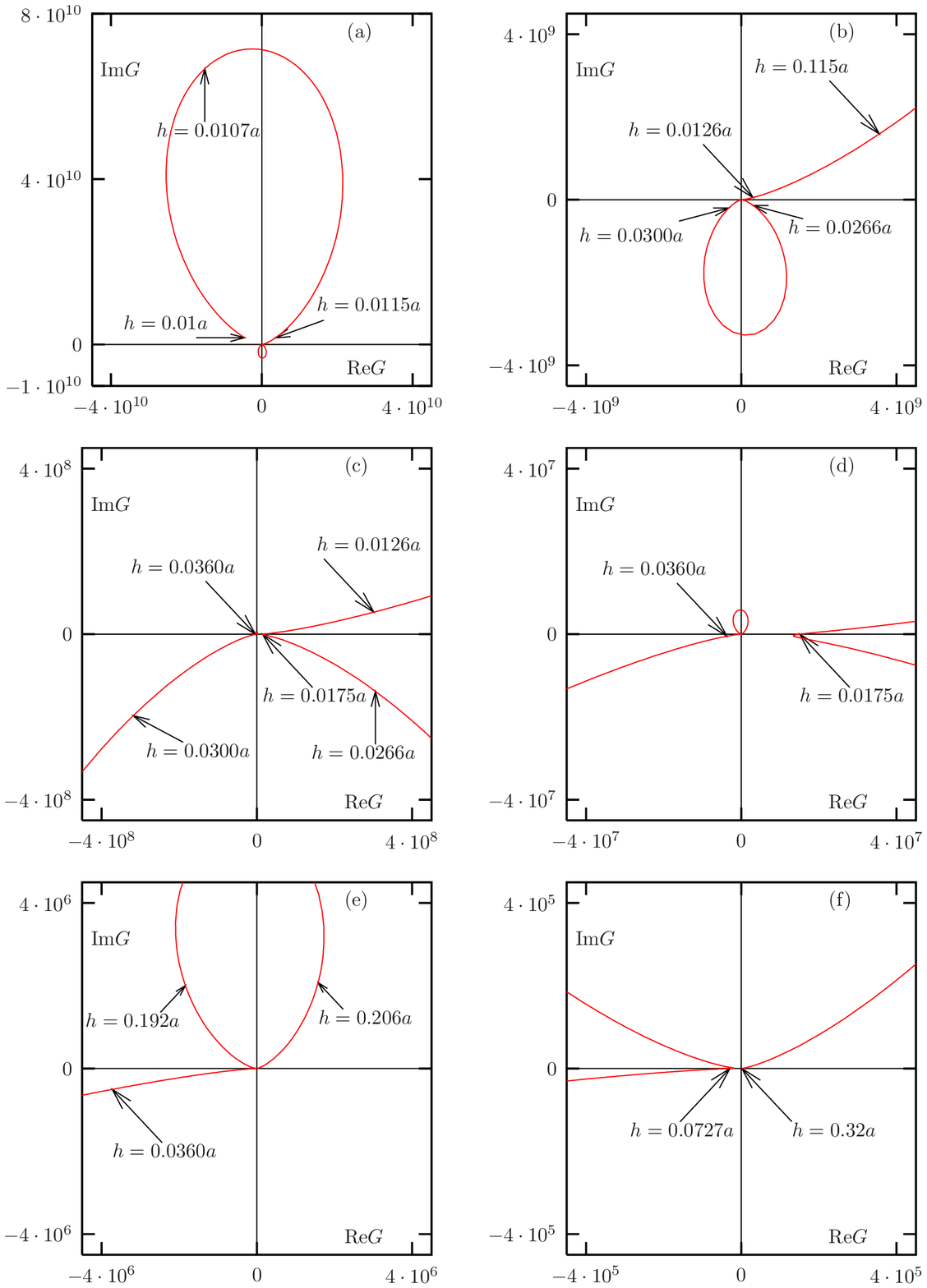,height=22cm,bbllx=50bp,bblly=0bp,bburx=600bp,bbury=700bp,clip=}}
%\centerline{\footnotesize \hspace*{-1.5cm}\input{eh_o=0.25_a.tex}\hspace*{-1.5cm}\input{eh_o=0.25_b.tex}}
%\centerline{\footnotesize \hspace*{-1.5cm}\input{eh_o=0.25_c.tex}\hspace*{-1.5cm}\input{eh_o=0.25_d.tex}}
%\centerline{\footnotesize \hspace*{-1.5cm}\input{eh_o=0.25_e.tex}\hspace*{-1.5cm}\input{eh_o=0.25_f.tex}}
\caption{\label{fig:eh_o=0.25} Same as in Fig.~\ref{fig:eh_o=0.20} but
  for $\omega=0.25\omega_p$.}
\end{figure}

\begin{figure}
\centerline{\psfig{file=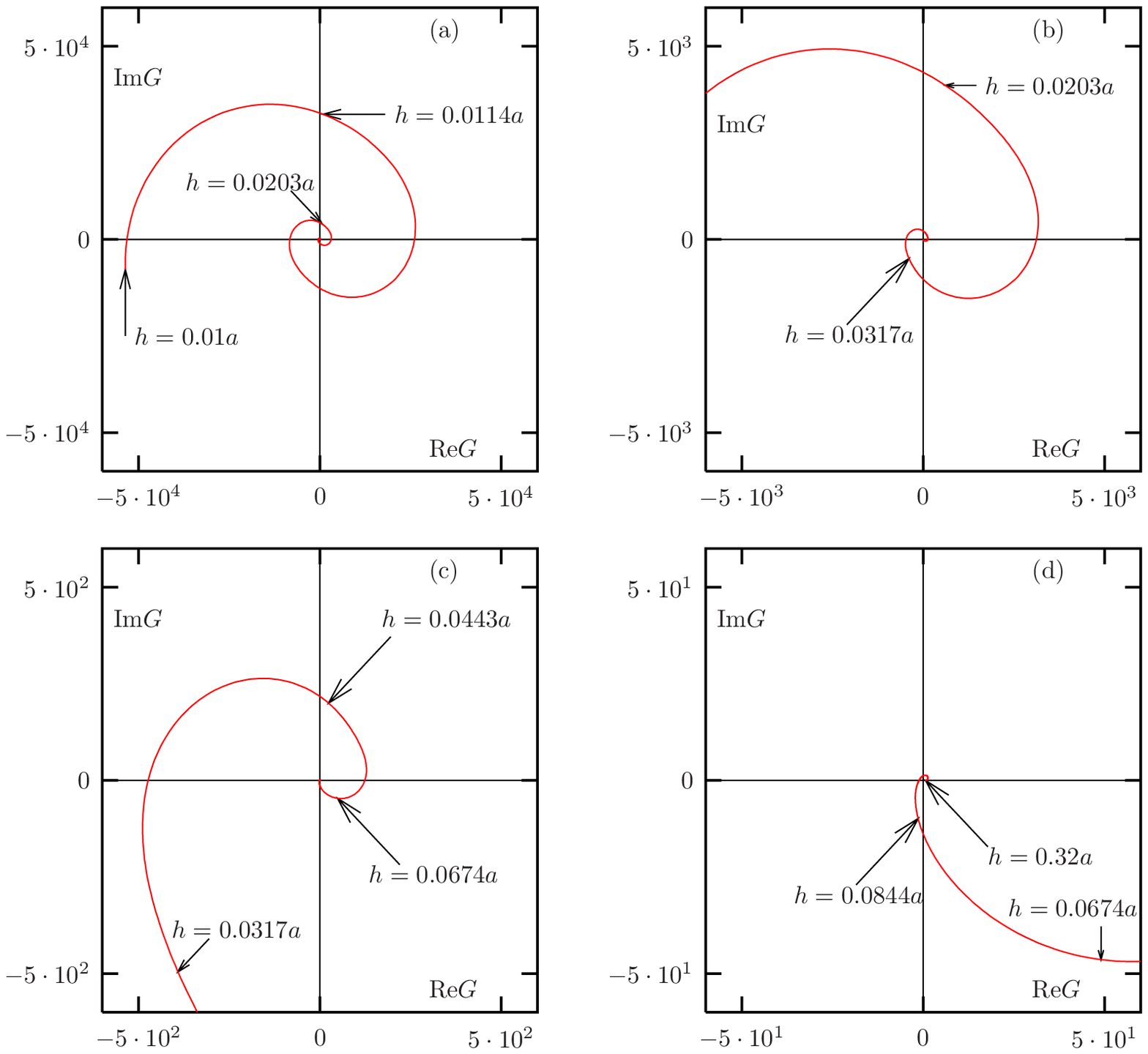,height=14.66cm,bbllx=50bp,bblly=250bp,bburx=600bp,bbury=700bp,clip=}}
%\centerline{\footnotesize \hspace*{-1.5cm}\input{eh_o=0.32_a.tex}\hspace*{-1.5cm}\input{eh_o=0.32_b.tex}}
%\centerline{\footnotesize \hspace*{-1.5cm}\input{eh_o=0.32_c.tex}\hspace*{-1.5cm}\input{eh_o=0.32_d.tex}}
\caption{\label{fig:eh_o=0.32} Same as in Fig.~\ref{fig:eh_o=0.20} but
  for $\omega=0.32\omega_p$.}
\end{figure}

In Figs.~\ref{fig:eh_o=0.20}-\ref{fig:eh_o=0.32}, we show the
parametric plots of the complex enhancement factor $G$ for the
following values of the ratio $\omega/\omega_p$: $0.20$, $0.25$ and
$0.32$. In the case $\omega/\omega_p=0.20$, the most dramatic change
of $G$ happens when $h$ changes from $0.36a$ to $0.41a$.  The phase of
$G$ changes in this interval of $h$ from $\approx \pi/4$ to $\approx
3\pi/4$. Overall, the phase of the enhancement factor can be tuned
from $\approx 0$ to $\approx \pi$ by tuning $h$ in the whole
considered interval.

Much more control over the phase of $G$ can be attained for
$\omega/\omega_p=0.25$, as shown in Fig.~\ref{fig:eh_o=0.25}. It is
interesting to note that the parametric curve shown in these figure is
approximately self-similar, consisting of several almost closed loops
which can be seen at different scales. The phase of $G$ changes in the
whole interval from $0$ to $2\pi$. Qualitatively similar curve was
also obtained for $\omega/\omega_p=0.30$ (data not shown).

Perhaps, the most interesting curve is obtained at
$\omega/\omega_p=0.32$ (Fig.~\ref{fig:eh_o=0.32}), although the
magnitude of $G$ is not as large for this value of $\omega/\omega_p$
as in Figs.~\ref{fig:eh_o=0.20}-\ref{fig:eh_o=0.25}.  The parametric
plot of $G$ is in this case a spiral. The phase of $G$ changes
monotonously from $\approx 0$ to $\approx 5\pi$. Thus, the curve makes
more than two full revolutions around the origin in the complex plane.
By varying the parameters in the Drude formula and the refractive
index of the host medium, it was found that the spiral shape of the
curve is typical when $\omega/\omega_p$ is close to the right bound of
the resonance interaction band (data not shown).

Thus, we have shown that by changing the inter-sphere separation $h$
from $0.01a$ to $0.32a$ it is possible to change the phase of the
enhancement factor $G$, and, consecutively, that of the effective
nonlinear polarizability $\alpha^{(3)}_{\rm eff}$ in its whole range.
Some limitations of the model used in this paper must be mentioned.
First, we did not account for direct electromagnetic coupling between
the nonlinearly polarizable molecule and the nanospheres, nor did we
take into account the nonlinearity of the metal itself. The latter
effect can be significant. Further, did not account for the fact that
the electric field in the gap is not constant but can change on the
scales comparable to molecular.  Although all these factors are
important if one seeks to calculate the nonlinear response of the
system with precision, inclusion of all these complications would make
the theoretical description intangible.  On the other hand, the
physical effect described in this paper does not originate due to any
of the approximations listed above. Instead, it is explained by the
resonant nature of the interaction between the electromagnetic field
and the nanosystem. When the spacing between the nanospheres is tuned,
different resonance modes are excited in the bisphere aggregate. This
results the change of the relative phase between the local field in
the gap $E_z$ and the external field $E_0$ and the characteristic
dependence of the enhancement factor $G$ on $h$ which is illustrated
in Figs.~\ref{fig:eh_o=0.20}-\ref{fig:eh_o=0.32}.  Since the resonance
nature of interaction is not altered by the effects mentioned above,
it is reasonable to expect that the fine tuning of the third-order
nonlinear response is achievable in nanosystems specifically
engineered for that purpose.

\section*{References}
\bibliography{abbrevplain,book,article}

\end{document}